\documentclass[12pt,preprint]{aastex}

\bibpunct{(}{)}{;}{a}{}{,}
\newcommand{\ignore}[1]{}
\newcommand{\eq}[1]{\begin{equation} #1 \end{equation}}
\newcommand{\eql}[2]{\begin{equation} \label{#1} #2 \end{equation}}

\newcommand{\rv}[1]{{#1}}
\newcommand{\rd}[1]{{#1}}
\newcommand{\rvv}[1]{{#1}}
\shorttitle{Energy Loss in Interstellar Bubbles}
\shortauthors{Krause \& Diehl}

\begin{document}

%% LaTeX will automatically break titles if they run longer than
%% one line. However, you may use \\ to force a line break if
%% you desire.

\title{\rd{Dynamics and }Energy Loss in Superbubbles}

%% Use \author, \affil, and the \and command to format
%% author and affiliation information.
%% Note that \email has replaced the old \authoremail command
%% from AASTeX v4.0. You can use \email to mark an email address
%% anywhere in the paper, not just in the front matter.
%% As in the title, use \\ to force line breaks.

\sloppy

\author{Martin
  G. H. Krause\altaffilmark{1,2} 
  and Roland Diehl\altaffilmark{2,1}}

\altaffiltext{1}{Excellence Cluster Universe, Boltzmannstr. 2, D-85748, Garching, Germany}
\altaffiltext{2}{Max-Planck-Institut f\"ur extraterrestrische Physik,
  Giessenbachstr.~1, 85741 Garching, Germany}
%\altaffiltext{3}{Geneva Observatory, University of Geneva, 51 Chemin des Maillettes, 1290 Versoix, Switzerland}
%% Mark off your abstract in the ``abstract'' environment. In the manuscript
%% style, abstract will output a Received/Accepted line after the
%% title and affiliation information. No date will appear since the author
%% does not have this information. The dates will be filled in by the
%% editorial office after submission.

\begin{abstract}
Interstellar bubbles \rd{appear to be smaller in observations than expected from calculations.} 
\rv{Instabilities at the shell boundaries create three-dimensional effects, and are probably responsible for part of this discrepancy.
We investigate instabilities and dynamics in superbubbles by 3D hydrodynamics simulations with time-resolved energy input from massive stars, including the supernova explosions.}
\rv{We find that the superbubble shells are accelerated by supernova explosions, coincident with substantial brightening in soft X-ray \rd{emission}. In between the explosions, the superbubbles lose energy efficiently\rvv{,} approaching the momentum-conserving snowplow limit.} 
\rd{This and enhanced radiative losses due to instabilities reduce the expansion compared to} the corresponding radiative bubbles \rv{in pressure-driven snowplow models with constant energy input.}
We note generally good agreement with observations of superbubbles \rvv{and some open issues. In particular, there are hints that the shell velocities in the X-ray-bright phases is underpredicted.}
%Intermittency of the energy input and instabilities are found to be an important factor for superbubble dynamics.
%Intermittency of the energy input and instabilities
%are found to be an important factor for superbubble dynamics.
\end{abstract}

%% Keywords should appear after the \end{abstract} command. The uncommented
%% example has been keyed in ApJ style. See the instructions to authors
%% for the journal to which you are submitting your paper to determine
%% what keyword punctuation is appropriate.


%% Authors who wish to have the most important objects in their paper
%% linked in the electronic edition to a data center may do so by tagging
%% their objects with \objectname{} or \object{}.  Each macro takes the
%% object name as its required argument. The optional, square-bracket 
%% argument should be used in cases where the data center identification
%% differs from what is to be printed in the paper.  The text appearing 
%% in curly braces is what will appear in print in the published paper. 
%% If the object name is recognized by the data centers, it will be linked
%% in the electronic edition to the object data available at the data centers   
%%
%% Note that for sources with brackets in their names, e.g. [WEG2004] 14h-090,
%% the brackets must be escaped with backslashes when used in the first
%% square-bracket argument, for instance, \object[\[WEG2004\] 14h-090]{90}).
%%  Otherwise, LaTeX will issue an error. 

\section{Introduction}\label{s:intro}
\rd{The interstellar medium is commonly seen to show %RD Bubbles, roundish 
cavities delimited by arcs and shells, }%RD , are ubiquitous 
%RD in the interstellar medium of 
in the Milky Way 
\citep[e.g.][]{Churchea06,Paladea12}
as well as in other star-forming galaxies \citep[e.g.][]{DPC01,Bagea11}.
\rd{Bubble centers seem to hold} %RD are found around 
individual stars and stellar explosion sites \citep[e.g.][]{Gruendea00,Green04}, as well as 
stellar groups and associations \citep[\rv{superbubbles}, e.g.][]{Cashea80,Maciea96,BdA06,Jaskea11}.
Interstellar bubbles are evidence of local energy input due to \rd{high-energy} %RD UV 
photons 
\citep[e.g.][]{OS55,DB11,KTK2014}
and/or mechanical energy from winds and explosions
\citep[e.g.][]{FHY03,vanVea09,Ntormea11,vanMea12,Georgea13,Krausea13a,RP13}.

The expansion of interstellar bubbles was \rd{explained} %RD described 
initially through 
self-similar, analytic \rd{descriptions} %RD models 
\rv{(e.g. \citeauthor{Sedov59} \citeyear{Sedov59}, point explosion;
\citeauthor{CMW75}\citeyear{CMW75}, \citeauthor{Weavea77}\citeyear{Weavea77}, 
constant power wind with and without cooling)}. Much effort has since then been devoted 
to incorporate additional complexity \rd{in such descriptions}, such as 
%radiative energy loss \citep[e.g.][]{Levy71,Weavea77}, 
non-uniform ambient density distributions 
\citep[e.g.][]{GSML95a} or mass-loading by entrained clouds \citep[e.g.][]{PDH01}, see
\citet{OMK88} for a review and more applications. 
%Motivated by increasingly detailed observations \citep[\rv{e.g.}][]{Gruendea00},
Structure \rd{of bubble interiors in terms} of density and composition %RD  in bubble interiors  
\citep[e.g.][]{vanVea09,Georgea13} as well as the shell structure
\citep[e.g.][]{Vish83,MLN93,Ntormea11,vMK12,Krausea13a,Pit13}
has been inferred from further analytical \rd{studies and } %RD work and
multi-dimensional simulations. 
%The interstellar bubble concept has also been applied beyond its original
%context to quasar winds and extragalactic radio sources 
%\citep[e.g.][]{GN73,Dysea80,Churea01,FL01,mypap03a,McNN07}, and is discussed 
%for the associated absorption lines in galaxy-scale outflows at high redshift 
%\citep[e.g.][]{Tapea04,Krause2005b,Verhea08,Chonea13,vGlea13}.

% save 27
%Radiative energy loss 
%occurs in most of these cases. It 
%becomes significant, 
%when the dynamical time 
%exceeds the cooling time in the shocked ambient gas. 
Observations of interstellar bubbles \rd{and constraints} %RD combined with observational constraints
on the\rd{ir} stellar content showed significant deviations from theory: 
the bubbles appear to be smaller 
and \rd{thus} to lose more energy \rd{during their evolution} than predicted in the models 
\citep[e.g.][]{Drisea95,GSML95a,Oey96,OG04,BB08,HCM09,Bruhwea10}. 
\rv{X-ray bright \rd{superbubbles} %RD ones 
are observed 
to have higher shell velocities than expected for their size and evolution time 
\citep[e.g.][]{Oey96}.}

Possible solutions to \rv{the first} \rd{discrepancy} %RD problem 
include 
a \rd{correction for the} underestimate of the ambient pressure \citep{OG04} and
blow-out \rv{in the case of superbubbles} \citep[e.g.][]{McLMcC88,BdA06}\rd{. The latter accounts for an observational bias, because} %RD any 
shells are best seen in 
the direction where
the density is the highest, and consequently \rd{where} the \rd{expansion is slowest.} %RD growth the slowest. 
\citet{HCM09} have suggested that energy may be lost from bubbles via 
leakage of hot gas through holes in the shell. Such holes \rd{have been found in simulations to} %RD do 
not appear easily, 
however, even in strong density inhomogeneities \citep{Pit13}.

\rv{
Time variability of the energy input has been suggested as an explanation for the
high velocity of X-ray-bright superbubbles. 
\citet{OG04} show in a 1D hydrodynamical simulation including radiative cooling
and heating due to photoionization that the supershells may 
be accelerated to the observed velocities after a supernova has exploded
in a bubble created beforehand by the massive-star winds.

%However, while the kinematics may be matched, the X-ray luminosity is difficult to explain:
%For example, the
%\rd{observed X-ray luminosity of superbubble DEM~L50 is significantly higher than predicted (as shown by \citet{Jaskea11}, using the model of
% \citet{OG04} taylored to match radius and velocity of the superbubble).}
%This might have been expected already
%from the calculations of \citet{CM90}, who showed that off-centre explosions are necessary to produce the observed luminosities 
%%RD , which is not possible
%\rd{(this cannot be described} in a 1D model, 
%compare also Sect.~\ref{sec:disc}).

Time variability has however also another effect, which is not captured 
in 1D models. It is well-known that shells are unstable to the Rayleigh-Taylor instability
whenever the shell accelerates, and to the Vishniac instability 
\citep{Vish83}, whenever it decelerates. The Vishniac instability leads to shell clumping, and thus quite possibly influences star formation. The Rayleigh-Taylor instability causes
filaments of shell gas to enter the superbubble, where it subsequently mixes with the 
bubble gas, thereby reducing its temperature and increasing its density.
We demonstrated both effects recently in 3D hydrodynamics simulations of superbubbles.
%RD In \citet[hereafter Paper~I]{Krausea13a} 
Even in the more complex case
of three close\rd{ly-spaced }%RD by 
massive stars, the filamentary network \rd{caused by} %RD of 
the Vishniac 
instability \rd{is} establishe\rd{d }%RD s
on 
%\rd{at} 
the shell early on,
%, during the wind phase of the most massive star\rd{,} 
and is present up to the end of the simulations  \citep[hereafter Paper~I]{Krausea13a} .
%RD  \citet{Jaskea11} using the model of
%RD \citet{OG04} taylored to match radius and velocity of the
%RD superbubble DEM~L50 showed that the predicted X-ray luminosity of this model is far 
%RD below the one observed for DEM~L50. 

%We found in  3D simulations that 
The instabilities lead
to an intermediate density region inside of the shell \citep[hereafter Paper~II]{Krausea14a}. 
The density is high enough
to produce an X-ray luminosity well comparable to \rd{the one seen in } X-ray-bright superbubbles.
%, and
%at the same time the temperature is generally well comparable to the observed ones. 
We calculated the spectrum in detail, assuming equilibrium ionisation\rd{,} %RD  structure 
and showed that it is \rvv{more complex than a thermal spectrum for a single temperature}. This \rd{is in good agreement with observations, as is }%RD , and 
the edge-brightened X-ray appearance. %RD morphology 
%RD in X-rays is in good agreement with observations.\\
%
Here, we show that time variability of the energy input also leads to a fundamental
change in bubble dynamics: for most of the time, superbubbles are in a 
{\em momentum-conserving} rather than a {\em pressure-driven} snowplow
phase. Additional energy loss occurs in the mixing region, because in certain
locations, densities and temperatures favorable for atomic line cooling are reached.
 We show in Sect.~\ref{sec:tsa} that also when taking this into account,
the momentum-conserving snowplow loses energy \rvv{most efficiently among the power law solutions, 
i.e. also faster} than the pressure-driven \rvv{snowplow.}
In Sect.~\ref{sec:comptosim} we demonstrate that our simulated superbubbles 
are almost always close to a momentum-conserving snowplow phase. They therefore
lose more energy and have smaller bubble radii than the classical pressure-driven 
snowplows. In Sect.~\ref{sec:kin-X} we 
show that the X-ray luminosity correlates strongly with shell acceleration,
%alleviating the problems in 1D models \citep[compare above]{Jaskea11}, 
and that 
a supershell may be accelerated beyond the velocity expected from even an 
adiabatic Weaver-model for short time intervals.
%, in similarly good agreement with the 
%observations as the 1D-models \citep[compare][]{Oey96}.
}

\section{%RD Energy 
\rv{Bubble dynamics} in the thin-shell approximation}\label{sec:tsa}
%\rv{Here we analyse a simple, fully integrable bubble model that allows for 
%arbitrary energy input and losses in both, shell and interior of the bubble.
%We show that also for this model the momentum-conserving snowplow is the power-law 
%solution with the strongest energy loss.}

Cooling in the shocked ambient medium of interstellar bubbles may lead to
the formation of a thin and cool shell \rv{(snowplow phase)}. 
Even when radiative energy losses are negligible, the strong decline of the density 
inwards of the outer shock in the classical solutions \citep[e.g.][]{Sedov59} justifies 
the thin-shell approximation for dynamical purposes \rv{\citep{Kompa60,BS95}}. 
The thin-shell model describes interstellar bubbles as systems of 
two homogeneous regions: the spherical interior carries the thermal energy,
$E_\mathrm{t}(t)$\rv{, which is subject to stellar energy input and adiabatic losses;
its pressure, $p(t)=(\gamma-1)E_\mathrm{t}/V(r)$, where $V(r)$ is the bubble volume,
regulates the dynamics of the surrounding shell
via the momentum equation
\eql{eq:m}{d(Mv)/dt = pA(r)\, .} 
Here, $Mv$ denotes the shell's momentum and $A(r)$
is the spherical surface area. The shell
carries the kinetic energy\rv{:} 
\eql{eq:e}{E_\mathrm{k}(t):=E(t)- E_\mathrm{t}(t)\, .}
This situation is also called {\em pressure-driven snowplow} 
\citep[their Sect.~VI.A]{OMK88},
 in contrast to the {\em momentum-driven snowplow}, where $p(t)=0$.
For constant energy injection (Weaver model), the pressure-driven snowplow 
dissipates a constant fraction of the input \rvv{power} at the leading 
radiative shock, and thus energy accumulates 
with time. The momentum-conserving snowplow loses energy at a rate
$\propto t^{-3/4}$ \citep[their Sect.~VI.A]{OMK88}.

\rd{We modified this description allowing for arbitrary energy loss in the leading shock, as well as the bubble interior by requiring only consistent dynamics regarding bubble pressure 
and shell acceleration, rather than summing up kinetic energy and adiabatic expansion energy loss \citep{mypap03a}. Therefore, } 
%RD A slight variation of this model has been introduced in \citep{mypap03a}. Instead of
%RD calculating the adiabatic expansion loss of the bubble interior explicitely
%RD and adding this energy
%RD to the shell's kinetic energy, it is only required that the dynamics is consistent, and 
%RD therefore, 
only eqs.~(\ref{eq:m}) and ~(\ref{eq:e}) are solved.  
 }

In \citet{mypap03a} we have shown that \rv{these equations} may be integrated
for arbitrary $E(t)$ and spherically
symmetric  density $\rho(r)$: 
\eq{\label{eq:bubeq}
  \int_0^r M(r^\prime) r^\prime\mathrm{d}r^\prime
  = 2 \int_0^t dt^\prime\int_0^{t^\prime} dt^{\prime\prime}
  E(t^{\prime\prime})\, ,}
where $M(r)=4 \pi \rho(r)\, r^3/3$. Inserting a constant energy \rvv{accumulation}
rate, $E(t)=Lt$, and a constant ambient density $\rho(r)=\rho_0$
recovers the classical solution by \citet{Weavea77}:
$ r=\alpha \left(L/\rho_0\right)^{1/5} t^{3/5}\, ,$
 but with $\alpha=0.83$.  This compares to $\alpha=0.76$ and
 $\alpha=0.88$\footnote{We actually calculate 0.82 from the expression
 they give.} in \citet{Weavea77} for radiative and non-radiative
shells, respectively. 
%Hence, radiation makes only a small difference
%in these {\em analytic} models, and we therefore adopt Equation~\ref{eq:bubeq} as
%generalisation of the \citet{Weavea77} model for an arbitrary energy
%input function. Equation~\ref{eq:bubeq} is also a generalisation in so
%far, as it does not rely on self-similarity.

For further analysis of the simulation results, we now discuss
suitable special solutions of equation~(\ref{eq:bubeq}).
For constant ambient density and an energy injection power law of 
$E(t)=c t^d$, the shell radius evaluates to:
\eq{
  r=\left(\frac{15c t^{d+2}}{2 \pi \rho_0 (d+1)(d+2)}\right)^{1/5}\, ,}
valid for $d>-1$. The shell velocity $v(t)$ follows by differentiation with
respect to $t$. From this we obtain the kinetic energy of the shell,
$E_\mathrm{k}(t) = 0.5 M(r(t)) v(t)^2$, and the
bubble's kinetic energy fraction:
\eql{eq:efrac}{
\epsilon_\mathrm{k}:=\frac{E_\mathrm{k}(t)}{E(t)} = \frac{d+2}{5(d+1)}\, . }

%
%-------------------------------------------------------------
%            Fig.: thin-shell approximation: energy fraction
%-------------------------------------------------------------
%                                          
\begin{figure}
 \plotone{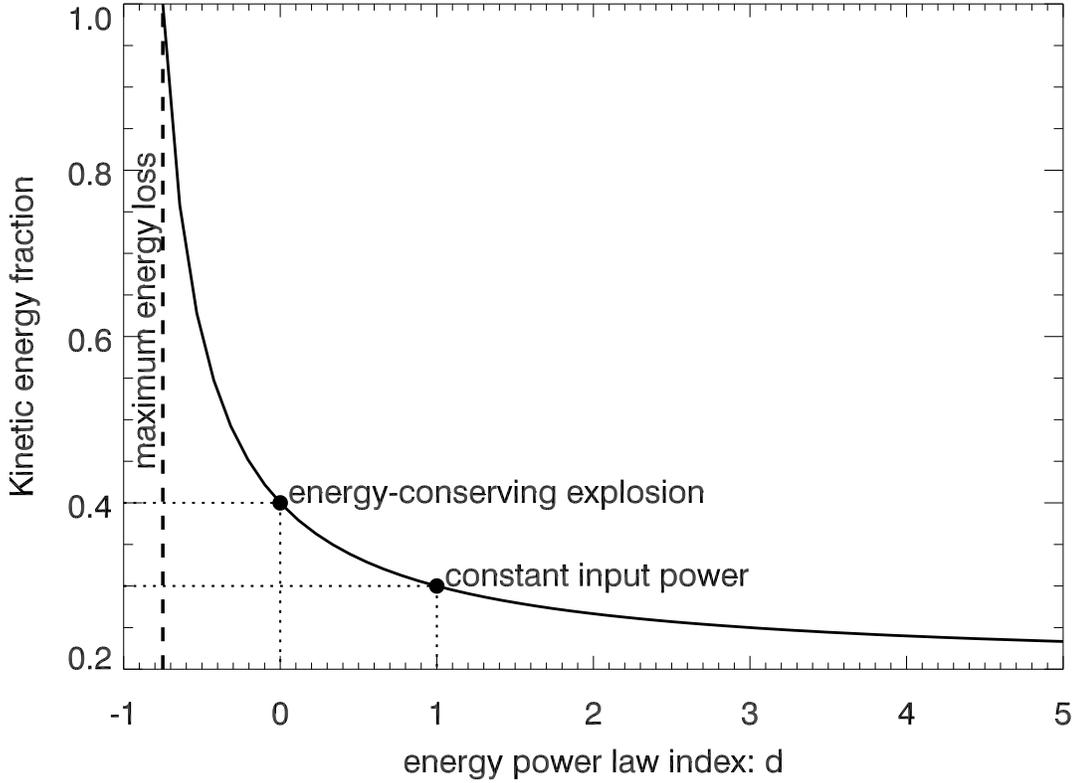}
\caption{Kinetic energy fraction in the thin-shell approximation, 
	$\epsilon_\mathrm{k}$ as a
  	function of the total energy evolution index $d$ (compare eq.~(\ref{eq:efrac}). 
	In the thin-shell approximation, $\epsilon_\mathrm{k}$ never falls below
	0.2 (horizontal axis). For a wind with constant input power ($d=1$) it is 
	$\epsilon_\mathrm{k}=0.3$; for an energy-conserving explosion ($d=0$), it is 
	$\epsilon_\mathrm{k}=0.4$. Both cases are marked in the plot 
	(dotted lines and filled circles).
	The lower limit for $d$ is $-3/4$ (dashed line), because the kinetic energy fraction 
	may not exceed unity. 
%Thus, within the thin-shell approximation, 
%	energy may not be lost faster than as $E(t)\propto t^{-3/4}$.
}
  \label{f:tsa-ekfrac}%
\end{figure}
%
%-------------------------------------------------------------
%    
Figure~\ref{f:tsa-ekfrac} shows the kinetic energy fraction as a function
of the power law index $d$ for the energy. Standard cases are marked: isolated
supernova, $\epsilon_\mathrm{k}(d=0)=0.4$, and constant luminosity wind, 
$\epsilon_\mathrm{k}(d=1)=0.3$. The total energy in a bubble 
may decay, e.g. due to radiation, and therefore $d$ may be negative.
The thin-shell approximation, however, restricts it to $d>-3/4$ (also marked in
Fig.~\ref{f:tsa-ekfrac}), because otherwise the total energy would be above 100~\%. 
\rv{This corresponds to the momentum-conserving snowplow.}
The lower limit to the kinetic energy fraction is~0.2, achieved
for a strong increase of the bubble energy.

\rv{This model is particularly useful for comparison to simulations, because it allows
for a direct comparison of the evolution of the bubble radius with $E(t)$ given 
by the stellar input to the corresponding one with $E(t)$ measured from the 
simulation, which takes into account the complex heating and cooling history due 
to the three-dimensional hydrodynamical evolution. }
%
%-------------------------------------------------------------
%            Fig.: 3S1-hr - energy tracks - detail
%-------------------------------------------------------------
%                                          
\begin{figure}
   \plotone{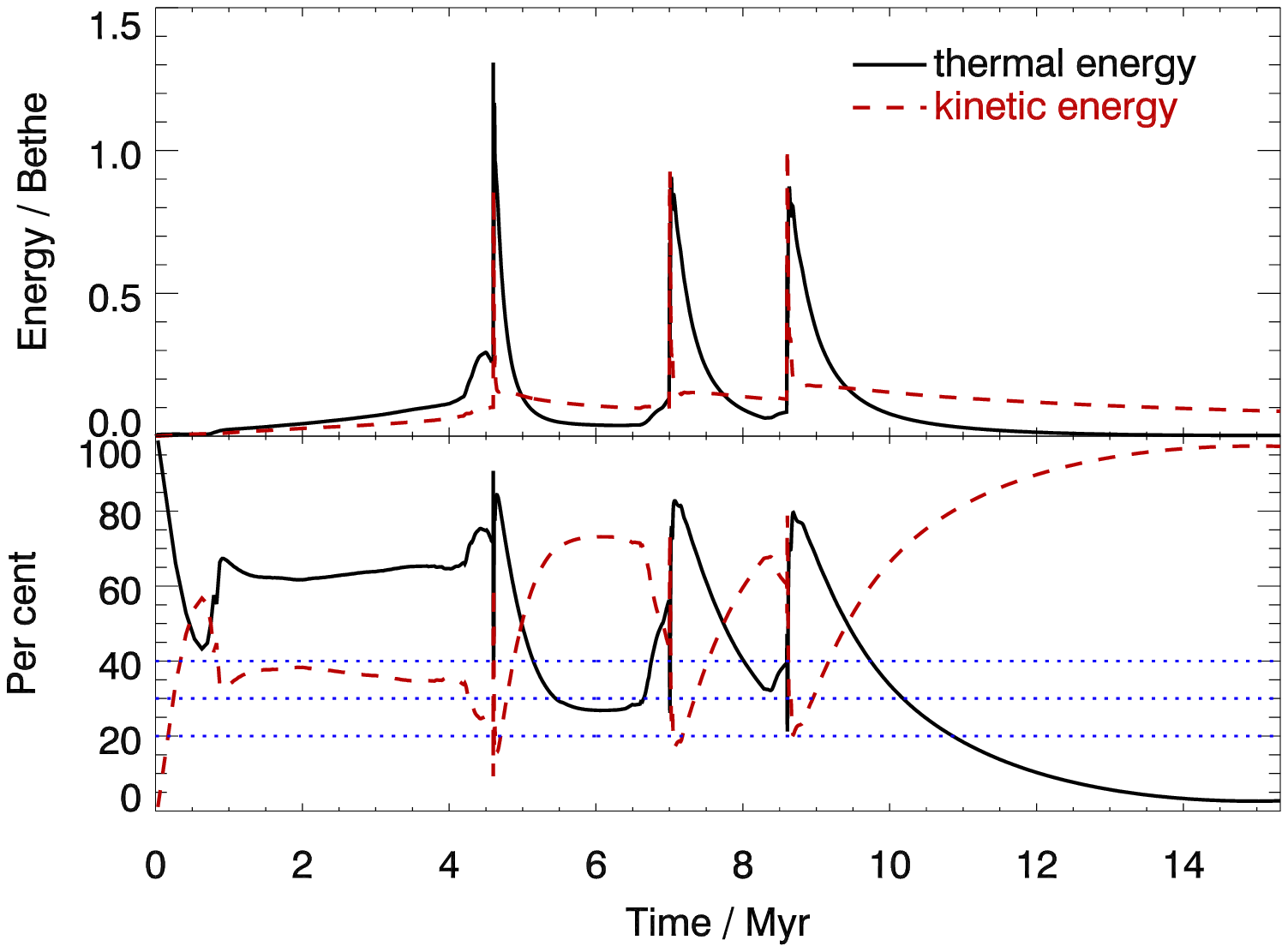}
  % \plotone{../../Wind_driven_ISM/idl/ekth_3S0-hr_69-72.eps}
\caption{Energy evolution for a 3D hydrodynamical simulation of the interstellar
	medium around three massive stars. 
	\rv{Top: Time evolution of the kinetic (red, dashed) and 
	thermal (black solid) energy. Bottom: corresponding 
	fractions of the total energy.} The dotted blue lines mark interesting values 
	for the kinetic energy 
	fraction derived in the thin-shell approximation, namely the lower limit 
	(20~\%), the case of a constant luminosity wind (30~\%) and the one of 
	the isolated\rv{, adiabatic} supernova (40~\%).}
  \label{f:3S1-hr_esplit}%
\end{figure}
%
%-------------------------------------------------------------
%    
\section{Comparison to 3D-hydrodynamics simulations}\label{sec:comptosim}

We now compare these analytic solutions
to the hydrodynamics simulations 
%of emerging superbubbles 
presented in \rv{Papers~I and~II}.
%RD For clarity,
\rvv{We show here a simulation with three massive stars (25, 32, and 60~$M_\odot$)
at tens of parsecs distance from each other,
where the individual bubbles merge early (3S1-hr in Paper~I)}. 
%RD The results are practically identical to the case with all three stars at the same location
%RD \rv{(Fig.12 in Paper~I)}. 

\begin{figure}
 \plotone{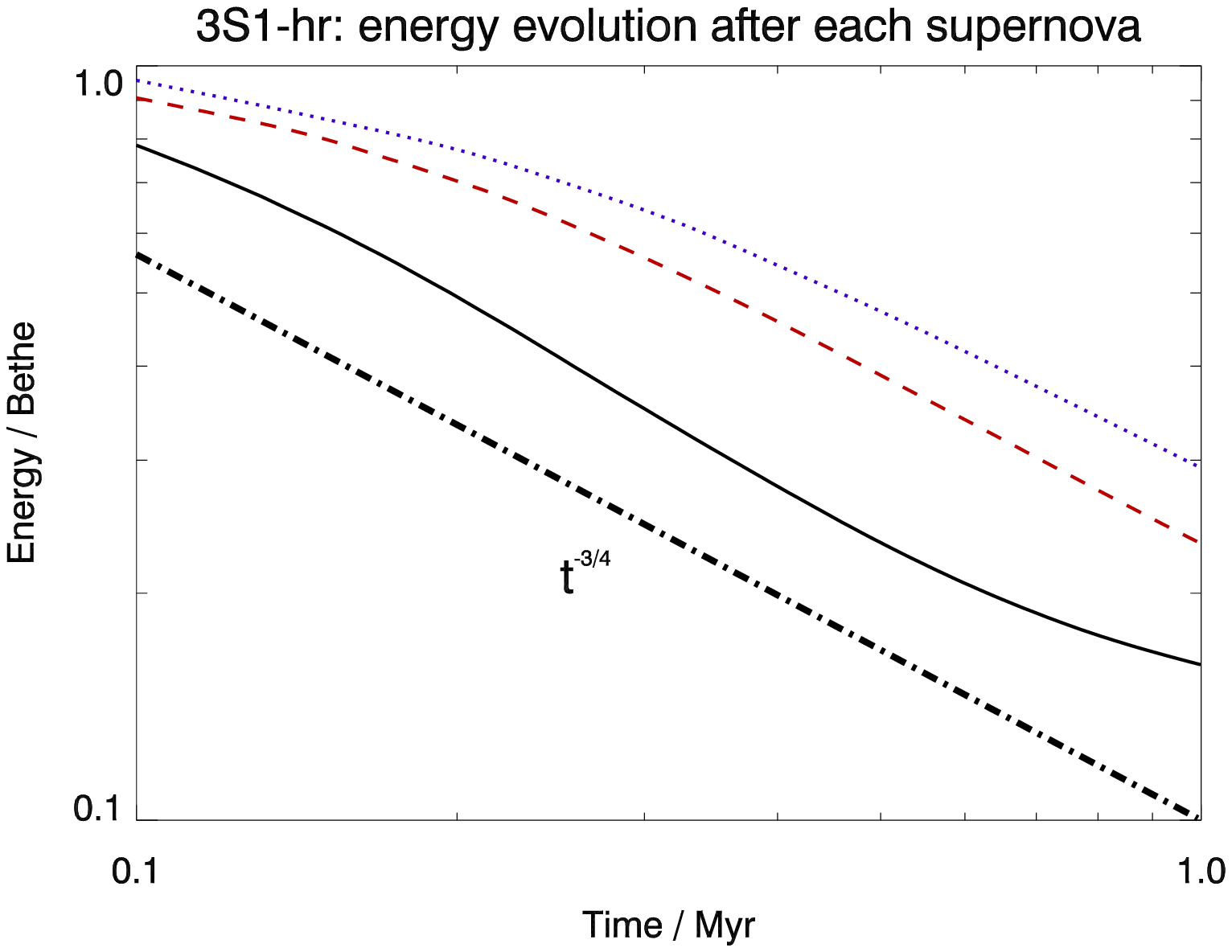}
\caption{Evolution of the total energy after each supernova, i.e. for
  1~Myr after the one at 4.6~Myr (solid black), at 7.0~Myr (dashed
  red) and at 8.6~Myr (dotted blue), respectively. The dash-dotted
  thick line is \rd{ the maximum energy
  loss rate which we derive for the thin-shell approximation, and is proportional to $t^{-3/4}$}. }
  \label{f:e-snplus}%
\end{figure}

\rv{The kinetic energy fraction varies in different phases of the superbubble evolution,
but is generally well described by the expectation from the thin-shell model in 
the respective phases (Fig.~\ref{f:3S1-hr_esplit}). In particular, in between supernovae
the kinetic energy fraction strongly dominates over the thermal one.}
%Clearly, the lower limit to the kinetic energy fraction in the thin-shell approximation
%(20~\%, compare above) is confirmed throughout the simulation, apart from the initial relaxation
%\rd{earlier than 1~Myr,} %RD before 1~Myr,
%which is due to the fact that we inject the energy purely in thermal form. Similarly, shortly %the energy evolution is dominated by the gas acceleration near the explosion site and not by the bubble dynamics.
%We find \rd{that the kinetic energy fraction settles at a value close }%RD an energy settling close
%to $\epsilon_\mathrm{k}=30$~\% for simulation times $1< t / \mathrm{Myr} <4$,
%where the energetics is dominated by the comparatively steady wind of the 60~$M_\odot$
%star, as expected from the thin-shell
%approximation. After each supernova, injected as thermal energy, the kinetic energy fraction first 
%oscillates due to the detailed hydrodynamical evolution inside the bubble, which is clearly %seen at high\rd{er} time resolution %\rd{simulations }
%(not shown here). After $\approx$50,000~yr, $\epsilon_\mathrm{k}$ has then adjusted, again very close
%to the 20~\% derived above for the limit of very strong energy increase.
%Subsequently, the kinetic energy fraction rises to very high values, as we expect for strong
%energy loss. We showed above that in this case, the total energy may not decrease more
%strongly with time $t$ than $t^{-3/4}$. 
We show the energy evolution for 1~Myr after each supernova for all
three supernovae on a log-log plot in Figure~\ref{f:e-snplus}. 
A power law \rd{behavior} with slope $-3/4$ is also indicated in
Figure~\ref{f:e-snplus}. Obviously, the energy loss of the bubble is %RD well 
limited by, and close to, 
the expectation from the thin-shell approximation. 
We note that the energy tracks of \citet[their Fig.~3, top left]{Thornea98} for the radiative phase
of an isolated supernova show the same effects as discussed here.

%
%-------------------------------------------------------------
%            Fig.: energy after sn
%-------------------------------------------------------------
%                                          
%
%-------------------------------------------------------------
%    
%
%-------------------------------------------------------------
%            Fig.: Radius over time
%-------------------------------------------------------------
%                                          
\begin{figure}
 \plotone{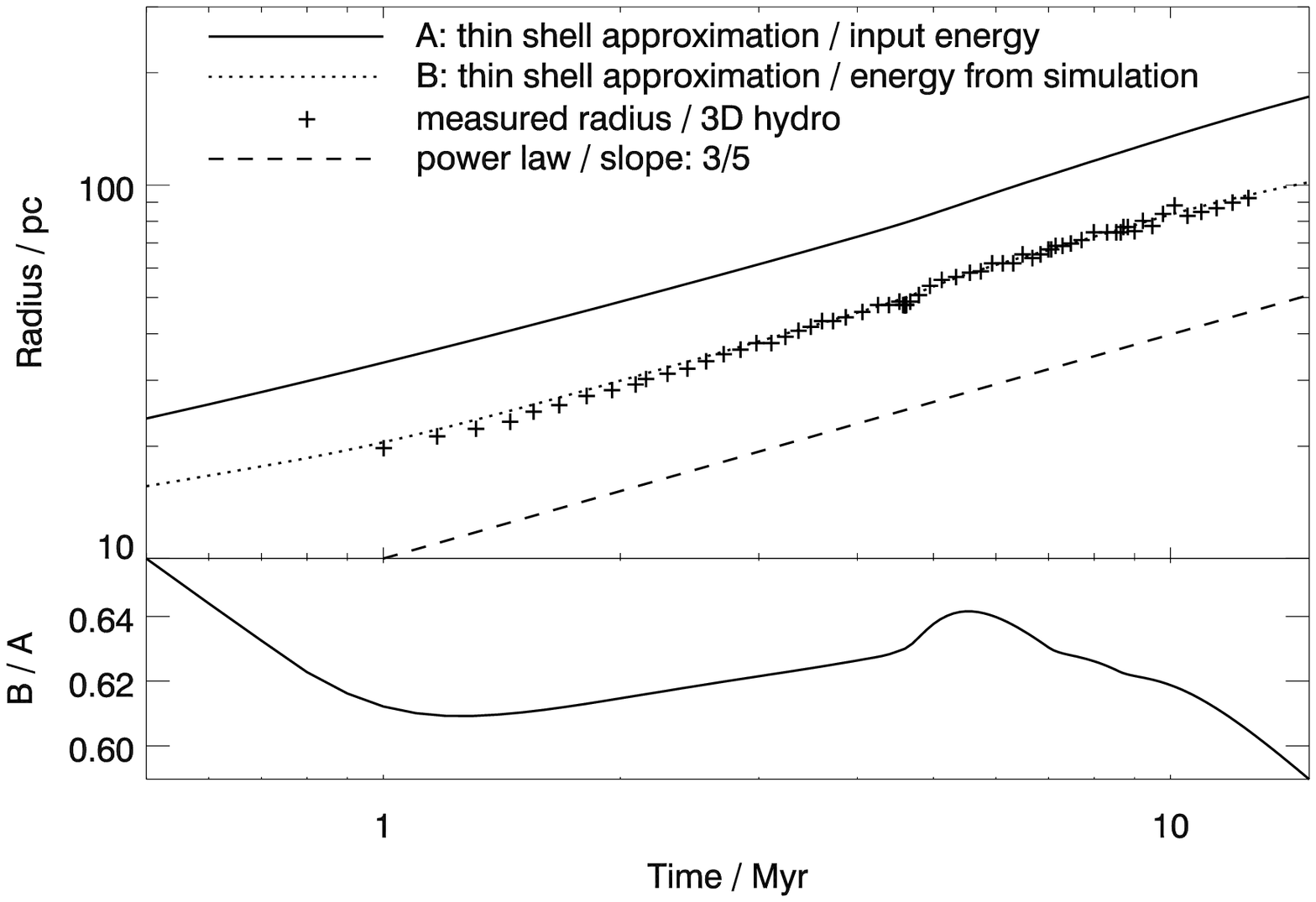}
\caption{\rd{Evolution of the bubble radius.} %RD Bubble radius over time. 
The top part compares the %RD radius expected from the
thin-shell approximation to the mean bubble
radius from the simulation ('+'-symbols) 
\rd{ in} a double-logarithmic scale.
%\rd{in two different treatments (see text) to our simulations}%RD, eq.~(\ref{eq:bubeq}),
 \rv{We show a thin-shell model using the full input energy (solid line, A),
%RD and, respectively, 
and one using the energy measured from the simulation (dotted line, B).}  %RD The dashed 
%RD line shows a power law with index $3/5$ for comparison.
The ratio of the two different thin-shell approximations, A/B, 
is shown in the bottom part. The radiative
energy loss reduces the bubble radius \rd{by 38~\% on average.}}%RD , averaged over time, by 38~\%.}
  \label{f:r-t}
\end{figure}
%
%-------------------------------------------------------------
%    

The thin-shell approximation also \rd{adequately reflects the link between total energy and shell radius} %RD connects the total energy in the simulation,
%RD well to the solid-angle-averaged radius of the bubble
 (Fig.~\ref{f:r-t}). The total 
energy differs from the input energy due to radiative losses.
While
the individual supernovae (4.6~Myr, 7~Myr, 8.6~Myr) are 
just discernible as mild steepenings of the curve, the
overall evolution with time $t$ is very well described by a $t^{3/5}$ power law,
as applicable for a bubble around an energy source with constant power.
\rv{The same conclusion was reached already by \citet{McLMcC88}. We discuss this 
in Sect.~\ref{sec:disc}, below.}
This supports the use of the $3/5$-law in analysis of observations in the literature
(compare Sect.~\ref{s:intro}).

However, the \rd{shell} radius is \rd{found now substantially smaller,} only $62\pm1.6$~\% of what it would be for an adiabatic bubble, i.e.
without radiative losses (Fig.~\ref{f:r-t}, bottom part). This compares to a reduction
by only 10~\% in the case of constant wind luminosity for the model of \citet{Weavea77}.
The smaller radius in our superbubble models from 3D hydrodynamics simulations
reflects the enhanced radiative energy losses compared to the \rv{Weaver-}model,
caused by the strong time-dependence of the energy input in realistic cases.

%=====================================================================
\rv{
\section{Correlation between shell acceleration and X-ray emission}\label{sec:kin-X}
%
%-------------------------------------------------------------
%            Fig.: shell acceleration with X-rays
%-------------------------------------------------------------
%                                          
\begin{figure}
% \plottwo{vtx_v1.eps}{vtx_3s1hr_v1.eps}
 \plotone{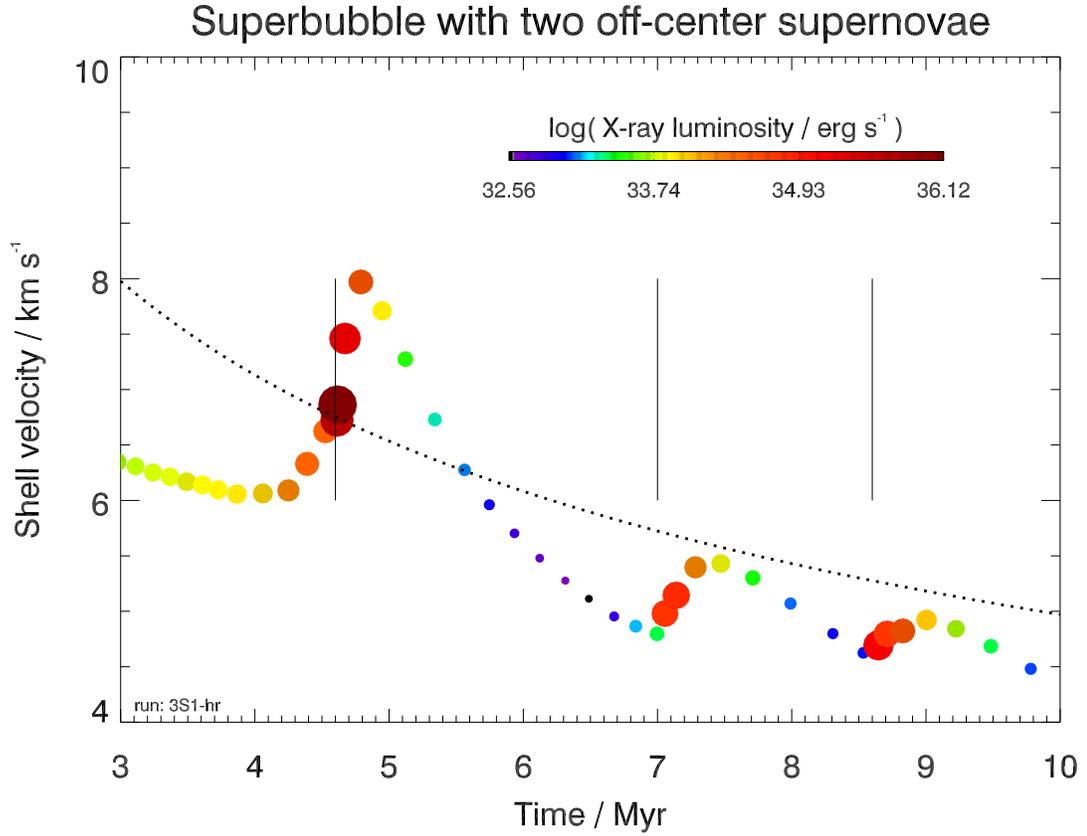}
\caption{\rd{Evolution of shell velocity.} %RD Shell velocity over time.
% for a case with off-center second and third supernovae. 
\rd{The s}ymbol size is \rd{chosen} proportional to the 
logarithm of the X-ray luminosity in the full ROSAT band (0.1-2.4~keV)\rd{; t}he color \rd{also} encodes \rd{X-ray luminosity.} %RD the same information according to the given color bar. Solid vertical lines indicate the times when supernovae explode. 
The dotted line gives the shell velocity for an adiabatic comparison model, where the energy input rate is constant in time and set to the average energy loss rate of the massive stars within 10~Myr. \rvv{Vertical lines indicate the times of the supernova explosions.}
%X-ray luminosity correlates strongly with shell acceleration, and the first 
%supernova pushes the shell beyond the velocity expected from the adiabatic model.
}
  \label{f:a-X}
\end{figure}
%
%-------------------------------------------------------------
%    
%X-ray-bright superbubbles have been found to have higher shell velocities than expected 
%from a constant energy input model (compare Sect.~\ref{s:intro}, above), which is well
%explained in 1D models employing a recent supernova \citet{OG04}. However, the
%X-ray luminosity from these 1D-models is far lower than expected and \citet{Jaskea11}
%explain this by unaccounted for thermal conduction. 
We \rd{calculate the superbubble X-ray emission spectrum using }%RD  have used 
a model for hot
plasma emission \citep['Mekal'][]{MGvdO85,MLvdO86,LOG95}. %RD  to calculate the X-ray 
%RD emission for our simulated superbubbles. 
General properties of the X-ray emission are discussed in Paper~II.
Entrainment and mixing lead to strong X-ray emission\rd{ in our simulations, }close to the observed luminosity.
\rd{The simulation results are converged in X-ray-bright phases.}
Figure~\ref{f:a-X} shows that the simulated X-ray luminosity strongly correlates with shell acceleration. The reason is that the supernova shock, which heats the intermediate density gas entrained due to 3D hydrodynamic instabilities, also accelerates the shell. As the shell accelerates, more instabilities entrain more cold gas\rd{, the mixing of which combines with adiabatic expansion to reduce the X-ray luminosity.} %RD. Mixing in this gas combined with the adiabatic expansion soon reduces the X-ray luminosity again. 
At the same time
the shell velocity starts to decline, quickly approaching the \rd{behavior in the }momentum conserving phase (compare Sect.~\ref{sec:comptosim}, above).
We find that the first supernova accelerates the shell beyond the velocity expected
from an adiabatic Weaver-model with constant, averaged energy input.
\section{Discussion}\label{sec:disc}
\subsection{Caveats}
%\rd{The approximate behavior of expanding shells in the interstellar medium is represented rather well in our simulations. However, we caution that we are still far from a realistic description. Here we discuss inadequacies in more detail.}
In \rd{the earliest stage of creation of an interstellar bubble, which corresponds to }the main-sequence phase of the most massive star, the shell always decelerates. 
Thus the contact surface between shocked wind and ambient gas 
is stable, and should therefore display a sharp change in density and temperature.
The finite resolution of \rd{simulations such as in }our code unavoidably produces intermediate 
temperature cells which artificially enhance radiative cooling beyond what one would expect 
from a thermal conduction model. Consequently, we discard this phase \rd{from further analysis, as in Paper~II, i.e.}, 
up to about 4~Myr of simulation time.%RD from further analysis, as in Paper~II. 

From the
first Wolf-Rayet phase onwards, the shell accelerates significantly (compare Fig.~\ref{f:a-X}). 
Consequently, Rayleigh-Taylor instabilities disturb the contact surface and lead to entrainment of shell gas into the bubble interior. The complex interplay
between Vishniac instability, Rayleigh-Taylor instability and mixing then determines the evolution.
The \rd{outcomes of these} processes are naturally resolution dependent, as, for example, the smallest
Rayleigh-Taylor mode grows fastest. Global properties may still converge 
with \rd{increasing} resolution, and thus be robust, if dominated by resolved modes.
%We have performed a resolution study for the representative simulation 3S1.
Radiative dissipation (Paper~I) and X-ray properties (Paper~II) %RD are 
\rd{have} essentially converged in this phase. 
Therefore, the results presented here should be robust.

Two reasons lead to enhanced energy loss in the simulated bubbles compared to
the standard solution by \citet{Weavea77}: First, \rd{radiation losses are larger} %RD there are enhanced radiative losses 
in a mixing layer of a few parsecs width. Densities are typically an order of magnitude higher than expected from thermal conduction, 
$\rho(x)=\rho_\mathrm{c} (1-x)^{-2/5}$ \citep{McLMcC88}, where 
$\rho_\mathrm{c}$ is the central density, and $x$ is the radial coordinate in units of the
shell radius. Second, the %RD simulated 
superbubbles are \rd{found in our simulations to be} almost always close to the momentum-conserving snowplow phase, which, as shown above, dissipates energy more efficiently than
the standard pressure-driven snowplow. Both processes contribute similarly
to \rd{increase} the overall energy loss. 

\subsection{Link to observations}
The momentum-conserving snowplow phases are a consequence of the time-dependent energy input, and therefore included in e.g. the 1D-models of \citet{Oey96} or \citet{OG04}. The latter authors also accounted for 
photoionization by massive stars. While the exact run of the density differs,
the effect of photoionization is quite comparable to mixing in creating an intermediate
density layer inside of the shell. \citet{Jaskea11} analysed the X-ray properties of two
superbubbles, including DEM~L50, using the code of \citet{OG04}. They found too low X-ray emission\rd{,} probably related to a very narrow
photoionized layer, and suggested \rvv{this was due to} unaccounted for thermal conduction. %RD In this case, 
\rd{W}e would expect that the mixing layer proposed
here can explain the X-ray emission \rd{at least equally well}. 
\rvv{We find a peak X-ray luminosity of $1.3\times10^{36}$~erg~s$^{-1}$, whereas \citet{Jaskea11} 
give $2-4.5\times10^{36}$~erg~s$^{-1}$. Similar X-ray-bright superbubbles in \citet{Oey96} have 
1.8 and $5.4\times10^{35}$~erg~s$^{-1}$ (DEM~L25 and DEM~L301, respectively). Within uncertainties, there is thus
reasonable agreement for the accelerating phase of our simulations,
up to about 0.2~Myr after a supernova. Independent confirmation of recent supernova activity is difficult. 
$^{26}$Al, radioactive with a half-life of 0.7~Myr and ejected by supernovae may provide an additional constraint in nearby ($\la$~kpc)
superbubbles. It is indeed detected \citep{Diehl02} in the high-velocity, X-ray-bright \citep{Oey96} Orion-Eridanus superbubble.}

For the superbubbles DEM~L25 and DEM~L301, in both of which the most massive star should have had around 60~$M_\odot$ which should have exploded a fraction of 
a Myr ago, \citet{OG04} find velocities of, respectively 40 and 
20~km/s in the
photoionised layer of their models. This compares to measurements of, respectively,
60 and 40~km/s from H$\alpha$, in broad agreement, given unaccounted for details of ionisation and geometry. Our simulated superbubbles 
have similar parameters to DEM~L25 and DEM~L301. At 4.5 Myr, their size is 
roughly 50~pc as the observed one. \citet{OG04} use an ambient density
of 17~cm$^{-3}$ to model these superbubbles, very comparable to the 
10~cm$^{-3}$ used in our simulations.  Still,
our simulated superbubbles reach only 8 km/s for the bulk shell velocity.
\rvv{This discrepancy might be due to neglect of photoionisation in our models, as the shells in 
\citet{OG04} are only partially ionised so that some low velocity parts would not contribute to the H-alpha emission.}

\citet{McLMcC88} showed that the shock waves of supernovae in superbubbles 
always turn subsonic well before they reach the supershell, because the thermal energy of the superbubble is larger than the energy increase due to the supernova.  In our 3D simulations, superbubbles cool much stronger and have only a fraction of a supernova energy immediately before each explosion. Consequently, the shock wave
reaches the shell even if the supernova is central. Off-centre supernovae,
which are mandatory to explain X-ray bright superbubbles in the standard framework
\citep{CM90} are no longer required in our 3D models, but produce a similar X-ray luminosity as central ones (Paper~II).

}

\section{Conclusions}\label{s:conc}

We have analysed the \rvv{dynamical evolution} of
emerging superbubbles in \rvv{3D simulations}\rd{,} and compared it to \rv{a} 
thin-shell 
\rv{model that allows for additional energy loss in the bubble interior}. 
We found 
that the latter describes the global aspects of the 3D  \rd{simulation} results very well.
%In particular, we found that t
\rvv{T}he total bubble energy 
\rv{in this thin-shell model} cannot decay faster with time $t$
than $t^{-3/4}$\rv{, as in the classical momentum-driven snowplow}, and that 
\rv{in the simulations} the cooling rate after energy injection by a supernova 
approaches this dynamical limit. 
\rv{Superbubbles are therefore much closer to a momentum conserving snowplow than to the pressure-driven
solution of the Weaver-model.}
 While the bubble radius \rd{in our simulations} still follows a $3/5$-law,
\rv{there is an enhanced energy loss due to the predominance of momentum-conserving phases and enhanced cooling in a mixing layer. The small bubble sizes \rd{that we find} are}
in \rvv{good} agreement with the observational \rd{results} %RD studies 
mentioned in Sect.~\ref{s:intro}.  
\rv{%RD While %RD our 
Shells \rd{in our simulations} accelerate beyond the velocity of a Weaver solution with equivalent, but time-averaged, power input \rd{from the stars and supernovae. This appears }coincident with X-ray brightening, as expected from observations.
\rvv{Our model does not explain the high H$\alpha$ velocities observed in X-ray-bright superbubbles, possibly due to lack of radiative transfer in the code.}}
%\rd{therefore the }shell velocities \rd{in our simulations} appear to be too low \rd{if compared to the Weaver model}. More detailed studies \rd{are required to settle these issues, and need to, e.g., take }
%%RD taking 
%into account photoionisation to determine H$\alpha$ luminosities \rd{, so that more observational constraints can be used to better approximate realistic shell evolution.}} %RDare required
%RD to investigate this issue.}
%Due to the nature of 
%our simulations, starting from isolated wind bubbles which merge into superbubbles, we believe that these 
%\rv{results} should be applicable to a wide range of interstellar bubbles. 

\acknowledgments

  We thank the referee for very useful suggestions that significantly improved the manuscript. This research was supported by the cluster of excellence ``Origin
  and Structure of the Universe'' (www.universe-cluster.de).

%\appendix

%\section{Appendix material}

\clearpage


\begin{thebibliography}{47}
\expandafter\ifx\csname natexlab\endcsname\relax\def\natexlab#1{#1}\fi

\bibitem[{{Bagetakos} {et~al.}(2011){Bagetakos}, {Brinks}, {Walter}, {de Blok},
  {Usero}, {Leroy}, {Rich}, \& {Kennicutt}}]{Bagea11}
{Bagetakos}, I., {Brinks}, E., {Walter}, F. et al. 2011, \aj, 141, 23

\bibitem[{{Bisnovatyi-Kogan} \& {Silich}(1995)}]{BS95}
{Bisnovatyi-Kogan}, G.~S. \& {Silich}, S.~A. 1995, Reviews of Modern Physics,
  67, 661

\bibitem[{{Breitschwerdt} \& {de Avillez}(2006)}]{BdA06}
{Breitschwerdt}, D. \& {de Avillez}, M.~A. 2006, \aap, 452, L1

\bibitem[{{Bruhweiler} {et~al.}(2010){Bruhweiler}, {Freire Ferrero}, {Bourdin},
  \& {Gull}}]{Bruhwea10}
{Bruhweiler}, F.~C., {Freire Ferrero}, R., {Bourdin}, M.~O., \& {Gull}, T.~R.
  2010, \apj, 719, 1872

\bibitem[{{Butt} \& {Bykov}(2008)}]{BB08}
{Butt}, Y.~M. \& {Bykov}, A.~M. 2008, \apjl, 677, L21

\bibitem[{{Cash} {et~al.}(1980){Cash}, {Charles}, {Bowyer}, {Walter},
  {Garmire}, \& {Riegler}}]{Cashea80}
{Cash}, W., {Charles}, P., {Bowyer}, S., et al. 1980, \apjl, 238, L71

\bibitem[{{Castor} {et~al.}(1975){Castor}, {McCray}, \& {Weaver}}]{CMW75}
{Castor}, J., {McCray}, R., \& {Weaver}, R. 1975, \apjl, 200, L107

\bibitem[{{Chu} \& {Mac Low}(1990)}]{CM90}
{Chu}, Y.-H. \& {Mac Low}, M.-M. 1990, \apj, 365, 510

\bibitem[{{Churchwell} {et~al.}(2006){Churchwell}, {Povich}, {Allen}, {Taylor},
  {Meade}, {Babler}, {Indebetouw}, {Watson}, {Whitney}, {Wolfire}, {Bania},
  {Benjamin}, {Clemens}, {Cohen}, {Cyganowski}, {Jackson}, {Kobulnicky},
  {Mathis}, {Mercer}, {Stolovy}, {Uzpen}, {Watson}, \& {Wolff}}]{Churchea06}
{Churchwell}, E., {Povich}, M.~S., {Allen}, D. et al. 2006, \apj, 649, 759

\bibitem[{{Dale} \& {Bonnell}(2011)}]{DB11}
{Dale}, J.~E. \& {Bonnell}, I. 2011, \mnras, 414, 321

\bibitem[{{Diehl}(2002)}]{Diehl02}
{Diehl}, R. 2002, \nar, 46, 547

\bibitem[{{Drissen} {et~al.}(1995){Drissen}, {Moffat}, {Walborn}, \&
  {Shara}}]{Drisea95}
{Drissen}, L., {Moffat}, A.~F.~J., {Walborn}, N.~R., \& {Shara}, M.~M. 1995,
  \aj, 110, 2235

\bibitem[{{Dunne} {et~al.}(2001){Dunne}, {Points}, \& {Chu}}]{DPC01}
{Dunne}, B.~C., {Points}, S.~D., \& {Chu}, Y.-H. 2001, \apjs, 136, 119

\bibitem[{{Freyer} {et~al.}(2003){Freyer}, {Hensler}, \& {Yorke}}]{FHY03}
{Freyer}, T., {Hensler}, G., \& {Yorke}, H.~W. 2003, \apj, 594, 888

\bibitem[{{Garc{\'{\i}}a-Segura} \& {Mac Low}(1995)}]{GSML95a}
{Garc{\'{\i}}a-Segura}, G. \& {Mac Low}, M.-M. 1995, \apj, 455, 145

\bibitem[{{Georgy} {et~al.}(2013){Georgy}, {Walder}, {Folini}, {Bykov},
  {Marcowith}, \& {Favre}}]{Georgea13}
{Georgy}, C., {Walder}, R., {Folini}, D. et al. 2013, \aap, 559, A69

\bibitem[{{Green}(2004)}]{Green04}
{Green}, D.~A. 2004, Bulletin of the Astronomical Society of India, 32, 335

\bibitem[{{Gruendl} {et~al.}(2000){Gruendl}, {Chu}, {Dunne}, \&
  {Points}}]{Gruendea00}
{Gruendl}, R.~A., {Chu}, Y.-H., {Dunne}, B.~C., \& {Points}, S.~D. 2000, \aj,
  120, 2670

\bibitem[{{Harper-Clark} \& {Murray}(2009)}]{HCM09}
{Harper-Clark}, E. \& {Murray}, N. 2009, \apj, 693, 1696

\bibitem[{{Jaskot} {et~al.}(2011){Jaskot}, {Strickland}, {Oey}, {Chu}, \&
  {Garc{\'{\i}}a-Segura}}]{Jaskea11}
{Jaskot}, A.~E., {Strickland}, D.~K., {Oey}, M.~S., {Chu}, Y.-H., \&
  {Garc{\'{\i}}a-Segura}, G. 2011, \apj, 729, 28

\bibitem[{{Kompaneets}(1960)}]{Kompa60}
{Kompaneets}, A.~S. 1960, Soviet Physics Doklady, 5, 46

\bibitem[{{Krasnobaev} {et~al.}(2014){Krasnobaev}, {Tagirova}, \&
  {Kotova}}]{KTK2014}
{Krasnobaev}, K.~V., {Tagirova}, R.~R., \& {Kotova}, G.~Y. 2014, \apj, 786, 90

\bibitem[{{Krause}(2003)}]{mypap03a}
{Krause}, M. 2003, \aap, 398, 113

\bibitem[{{Krause} {et~al.}(2014){Krause}, {Diehl}, {B{\"o}hringer},
  {Freyberg}, \& {Lubos}}]{Krausea14a}
{Krause}, M., {Diehl}, R., {B{\"o}hringer}, H., {Freyberg}, M., \& {Lubos}, D.
  2014, \aap, 566, A94

\bibitem[{{Krause} {et~al.}(2013){Krause}, {Fierlinger}, {Diehl}, {Burkert},
  {Voss}, \& {Ziegler}}]{Krausea13a}
{Krause}, M., {Fierlinger}, K., {Diehl}, R. et al. 2013, \aap, 550, A49

\bibitem[{{Liedahl} {et~al.}(1995){Liedahl}, {Osterheld}, \&
  {Goldstein}}]{LOG95}
{Liedahl}, D.~A., {Osterheld}, A.~L., \& {Goldstein}, W.~H. 1995, \apjl, 438,
  L115

\bibitem[{{Mac Low} \& {McCray}(1988)}]{McLMcC88}
{Mac Low}, M.-M. \& {McCray}, R. 1988, \apj, 324, 776

\bibitem[{{Mac Low} \& {Norman}(1993)}]{MLN93}
{Mac Low}, M.-M. \& {Norman}, M.~L. 1993, \apj, 407, 207

\bibitem[{{Maciejewski} {et~al.}(1996){Maciejewski}, {Murphy}, {Lockman}, \&
  {Savage}}]{Maciea96}
{Maciejewski}, W., {Murphy}, E.~M., {Lockman}, F.~J., \& {Savage}, B.~D. 1996,
  \apj, 469, 238

\bibitem[{{Mewe} {et~al.}(1985){Mewe}, {Gronenschild}, \& {van den
  Oord}}]{MGvdO85}
{Mewe}, R., {Gronenschild}, E.~H.~B.~M., \& {van den Oord}, G.~H.~J. 1985,
  \aaps, 62, 197

\bibitem[{{Mewe} {et~al.}(1986){Mewe}, {Lemen}, \& {van den Oord}}]{MLvdO86}
{Mewe}, R., {Lemen}, J.~R., \& {van den Oord}, G.~H.~J. 1986, \aaps, 65, 511

\bibitem[{{Ntormousi} {et~al.}(2011){Ntormousi}, {Burkert}, {Fierlinger}, \&
  {Heitsch}}]{Ntormea11}
{Ntormousi}, E., {Burkert}, A., {Fierlinger}, K., \& {Heitsch}, F. 2011, \apj,
  731, 13

\bibitem[{{Oey}(1996)}]{Oey96}
{Oey}, M.~S. 1996, \apj, 467, 666

\bibitem[{{Oey} \& {Garc{\'{\i}}a-Segura}(2004)}]{OG04}
{Oey}, M.~S. \& {Garc{\'{\i}}a-Segura}, G. 2004, \apj, 613, 302

\bibitem[{{Oort} \& {Spitzer}(1955)}]{OS55}
{Oort}, J.~H. \& {Spitzer}, Jr., L. 1955, \apj, 121, 6

\bibitem[{{Ostriker} \& {McKee}(1988)}]{OMK88}
{Ostriker}, J.~P. \& {McKee}, C.~F. 1988, Reviews of Modern Physics, 60, 1

\bibitem[{{Paladini} {et~al.}(2012){Paladini}, {Umana}, {Veneziani},
  {Noriega-Crespo}, {Anderson}, {Piacentini}, {Pinheiro Gon{\c c}alves},
  {Paradis}, {Tibbs}, {Bernard}, \& {Natoli}}]{Paladea12}
{Paladini}, R., {Umana}, G., {Veneziani}, M. et al. 2012, \apj, 760, 149

\bibitem[{{Pittard}(2013)}]{Pit13}
{Pittard}, J.~M. 2013, \mnras, 435, 3600

\bibitem[{{Pittard} {et~al.}(2001){Pittard}, {Dyson}, \& {Hartquist}}]{PDH01}
{Pittard}, J.~M., {Dyson}, J.~E., \& {Hartquist}, T.~W. 2001, \aap, 367, 1000

\bibitem[{{Rogers} \& {Pittard}(2013)}]{RP13}
{Rogers}, H. \& {Pittard}, J.~M. 2013, \mnras, 431, 1337

\bibitem[{{Sedov}(1959)}]{Sedov59}
{Sedov}, L.~I. 1959, {Similarity and Dimensional Methods in
  Mechanics}, Academic Press,
New York

\bibitem[{{Thornton} {et~al.}(1998){Thornton}, {Gaudlitz}, {Janka}, \&
  {Steinmetz}}]{Thornea98}
{Thornton}, K., {Gaudlitz}, M., {Janka}, H.-T., \& {Steinmetz}, M. 1998, \apj,
  500, 95

\bibitem[{{van Marle} \& {Keppens}(2012)}]{vMK12}
{van Marle}, A.~J. \& {Keppens}, R. 2012, \aap, 547, A3

\bibitem[{{van Marle} {et~al.}(2012){van Marle}, {Meliani}, \&
  {Marcowith}}]{vanMea12}
{van Marle}, A.~J., {Meliani}, Z., \& {Marcowith}, A. 2012, \aap, 541, L8

\bibitem[{{van Veelen} {et~al.}(2009){van Veelen}, {Langer}, {Vink},
  {Garc{\'{\i}}a-Segura}, \& {van Marle}}]{vanVea09}
{van Veelen}, B., {Langer}, N., {Vink}, J., {Garc{\'{\i}}a-Segura}, G., \& {van
  Marle}, A.~J. 2009, \aap, 503, 495

\bibitem[{{Vishniac}(1983)}]{Vish83}
{Vishniac}, E.~T. 1983, \apj, 274, 152

\bibitem[{{Weaver} {et~al.}(1977){Weaver}, {McCray}, {Castor}, {Shapiro}, \&
  {Moore}}]{Weavea77}
{Weaver}, R., {McCray}, R., {Castor}, J., {Shapiro}, P., \& {Moore}, R. 1977,
  \apj, 218, 377

\end{thebibliography}
\end{document}